\documentclass{elsart5p}

\usepackage{natbib}
\usepackage{psfig}
\usepackage{graphics}

\begin{document}

\begin{frontmatter}

\title{Supernova remnants with magnetars: clues to magnetar formation}
\author{Jacco Vink}
\ead{j.vink@astro.uu.nl}
\address{
  Astronomical Institute Utrecht, University Utrecht,
  P.O. Box 80000, 3508 TA, Utrecht, The Netherlands}

\begin{abstract}
In this paper I discuss the lack of observational evidence that magnetars
are formed as rapidly rotating neutron stars.
Supernova remnants containing magnetars do not show the excess of kinetic
energy expected for such a formation scenario, nor is there any
evidence for a relic pulsar wind nebula.
However, it could be that magnetars are formed with somewhat slower
rotation periods, or that not all excess rotational energy was used to
boost the explosion energy, for example as a result of
gravitational radiation.  Another observational tests for the
rapid initial period hypothesis is to look for statistical evidence
that about 1\% of the observed supernovae have an additional
 $10^{40}-10^{44}$~erg/s excess energy during the first year, 
caused by the spin down luminosity of a magnetar. 

An alternative scenario for the high magnetic fields of magnetars is the 
fossil field hypothesis, in which the magnetic field is inherited from the
progenitor star. Direct observational tests for this hypothesis are harder
to formulate, unless the neutron star formed in the SN1987A explosion
emerges as a slowly rotating magnetar.

Finally, I
point out the possible connection between the jets in Cas A and its X-ray 
point source: the jets in Cas A may indicate that the explosion was
accompanied by an X-ray flash, probably powered by a rapidly rotating
compact object. However, the point source in Cas A does not
seem to be a rapidly rotating neutron star. This
suggests that Cas A contains a  neutron star that has 
slowed down considerably in 330~yr,
requiring a dipole magnetic field of $B>5\times 10^{13}$~G.
The present day lack of evidence for a relic radio pulsar wind nebula
may be used to infer an even higher magnetic field of $10^{15}$~G.
\end{abstract}

\end{frontmatter}

\newcommand{\chandra}{{\em Chandra}}
\newcommand\asca{{\em ASCA}}
\newcommand\xmm{{\em XMM-Newton}}
\newcommand\sax{{\em BeppoSAX}}
\newcommand\rxte{{\em RXTE}}
\newcommand\msun{{$M_{\odot}$}}
\newcommand{\kms}{{km\,s$^{-1}$}}
\newcommand{\net}{{$n_{\rmn e}t$}}

\newcommand\spex{{SPEX}}
\newcommand\xspec{{XSPEC}}

\newcommand{\adspr}{{AdSpR\ }}
\newcommand{\apj}{{ApJ\ }}
\newcommand{\apjs}{{ApJS\ }}
\newcommand{\apjl}{{ApJ\ }}
\newcommand{\aj}{{AJ\ }}
\newcommand{\aap}{{A\&A\ }}
\newcommand{\aaps}{{A\&AS\ }}
\newcommand{\nat}{{Nat\ }}
\newcommand{\jetp}{{JETP\ }}
\newcommand{\mnras}{{MNRAS\ }}
\newcommand{\phrvl}{{PhRvL\ }}
\newcommand{\phrc}{{PhRvC\ }}
\newcommand{\prc}{{PhRvC\ }}
\newcommand{\araa}{{ARA\&A\ }}
\newcommand{\pasj}{{PASJ\ }}
\newcommand{\pasp}{{PASP\ }}
\newcommand{\npa}{{NuPhA\ }}
\newcommand{\iaucirc}{{IAU circ.\ }} 
\newcommand{\aplett}{{Astrophysical Letters\ }} 
\newcommand{\gca}{Geochimica et Cosmochimica Acta\ }

\def\lesssim{\mathrel{\hbox{\rlap{\hbox{\lower4pt\hbox{$\sim$}}}\hbox{$<$}}}}
\def\gtrsim{\mathrel{\hbox{\rlap{\hbox{\lower4pt\hbox{$\sim$}}}\hbox{$>$}}}}
\def\la{\mathrel{\hbox{\rlap{\hbox{\lower4pt\hbox{$\sim$}}}\hbox{$<$}}}}
\def\ga{\mathrel{\hbox{\rlap{\hbox{\lower4pt\hbox{$\sim$}}}\hbox{$>$}}}}

\section{Introduction}

\begin{figure*}
\centerline{
  \hskip 0.05\textwidth
  \psfig{figure=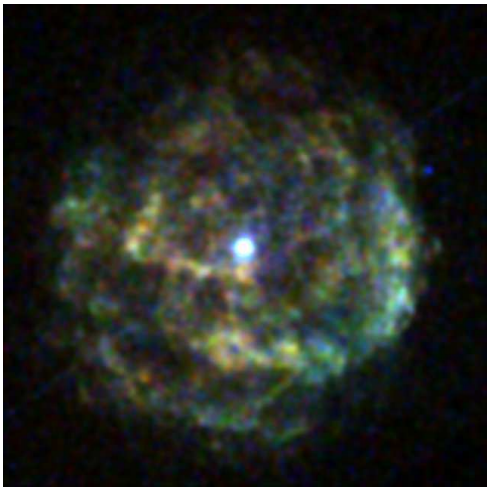,width=0.35\textwidth}
  \hskip 0.1\textwidth
  \psfig{figure=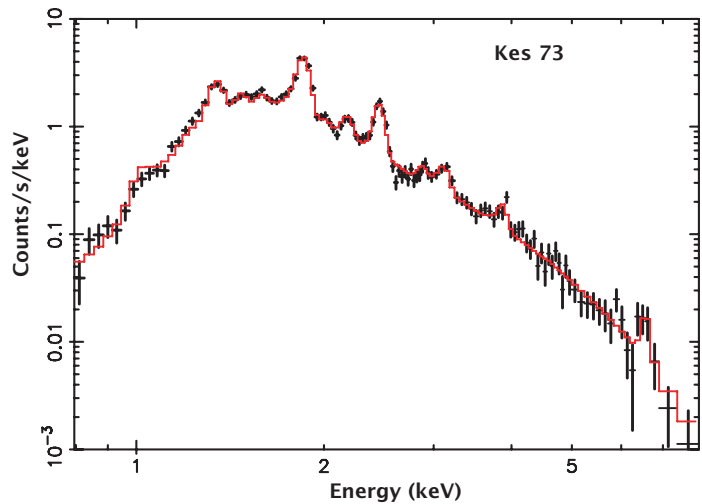,width=0.5\textwidth}
}
\caption{Left: Chandra X-ray image of Kes 73. The AXP 1E1841-045 is
the bright point in the center (saturated in this image in order to
bring out the SNR). Right: The \xmm\ EPIC X-ray spectrum
of Kes 73, excluding the AXP. The spectral model consists of
a two temperature non-equilibrium ionization model, provided by the SPEX
spectral code \citep{kaastra00}.}
\label{fig-kes73}       
\end{figure*}
Over the last decade ``Anomalous X-ray Pulsars'' (AXPs) and
``Soft Gamma Ray Repeaters'' (SGRs) 
have become one of the most exciting topics in 
neutron star research \citep[][for a review]{woods04}.
The phenomenology of AXPs/SGRs, such as their bursting
behavior and their period clustering between $P = 5 -12$~s 
have been explained by a variety of mundane and exotic\footnote{Both adjectives
are not meant to be derogatory; simple, mundane theories
are appealing to most scientists, and what is considered
exotic changes with time: neutron stars and black holes
were once considered very exotic topics.}
models, including fall-back disks 
\citep[e.g.][]{chatterjee00,alpar01}
and the idea that AXPs/SGRs are quarkstars \citep{ouyed06,niebergal06}.
However, the most widely accepted model for AXPs/SGRs is that they are
magnetars \citep{duncan92}, i.e. neutron stars with ultra high magnetic
surface magnetic fields ($10^{14}-10^{15}$~G).

The theoretical ideas about magnetars and how they are created were
formulated in a series of papers by Duncan and Thompson
\citep{duncan92,thompson93,duncan96}.
They suggested that magnetars form through magnetic field amplification
by a powerful dynamo, which was active during the,
highly convective, proto-neutron star phase. For this
dynamo to operate an initial period is needed that is shorter than the
typical convective overturn time of $\sim 3$~ms \citep{duncan96}.
The time scale in which a magnetar field is formed should be of 
the order of 10~s.

In this paper I will assume that AXPs/SGRs are indeed magnetars. However,
I will discuss whether the most widely accepted theory as to why
some neutron stars have very high magnetic fields, namely
due to a convective dynamo in a extremely rapidly rotating proto-neutron star,
is supported by observational data. Recent observational
results have raised some doubt on this canonical magnetar formation
theory  \citep{ferrario06,vink06c}.\footnote{Note that the
quark-star model of \citet{niebergal06} also assumes magnetar-like
fields for AXPs/SGRs, but assuming initial periods of 5~s.
For the fall-back disk model there is no need to invoke high magnetic
fields, although hybrid models, a fall-back disk and magnetar-like
fields, have also been considered \citep{ertan07}.
}

An alternative theory for magnetar formation is flux conservation,
often referred to as the fossil field hypothesis:
The high magnetic fields of magnetars directly reflect the high magnetic
fields of the cores of their progenitors \citep{woltjer64}.
In this case magnetar magnetic fields represent 
the tail of the magnetic field distribution of neutron stars.\footnote{
Recently two new theories about
magnetar formation have been proposed.
\citet{geppert06}  proposed yet another scenario, 
in which all neutron stars are born with $\sim 10^{15}$~G fields,
but the magnetic field is only stable in those neutron stars that spin faster 
than $\sim 6$~ms.\citet{bhattacharya07} suggests that magnetar
fields are created when a phase transition to exotic, magnetized matter
occurs in the most massive neutron stars. 
This model makes the need for a high initial spin period obsolete.
}

Interestingly, the two hypotheses have opposite implication for
the angular momenta of the progenitor stars. The hypothesis
of \citet{duncan92} implies that magnetars are formed from rapidly rotating
stellar cores, whereas the fossil field hypothesis 
implies large stellar magnetic fields, which result in an effective
rotational coupling of the stellar core with the stellar surface. In that case
stellar winds remove angular momentum from the core, resulting in more slowly
rotating stellar cores \citep{spruit02,heger05}.

In that light the debate over magnetic fields and rotation of magnetars
is interesting for neutron star initial rotation periods in general. 
For example the
results of \citet{ott06} show that short initial pulsar periods
are a natural result of core collapse supernovae, only if one ignores
the progenitor's magnetic field \citep[see also][]{heger05}.

Here I will discuss the clues that supernova remnants (SNRs) 
containing magnetars (AXPs, SGRs) provide us about magnetar formation.
This is partially based on the results presented in \citet{vink06c}. However,
I will also discuss the X-ray point source in the SNR Cas A, a magnetar
candidate. As I will discuss the presence of a putative magnetar 
in a SNR with a jet/counter jet structure may be an indication that the
neutron star was born rapidly rotating.

\section{Supernova remnants and the case for slowly rotating 
proto-neutron stars}

\subsection{The supernova remnant/magnetar connection}
AXPs and SGRs appear to be very young neutron stars of
typically a few thousand years old, as indicated by their spin-down
ages. The timing properties of AXPs and SGRs 
are, however, somewhat erratic, 
so that characteristic 
ages  are even less reliable age estimators than for other young pulsars.
However, the young ages are supported by the fact that of the 12 AXPs and SGRs
listed by \citet{woods04}
four are associated with SNRs
\citep[see ][for a discussion on SNR/magnetar associations]{gaensler04}.

The presence of AXPs and SGRs make these SNRs interesting in light of
the magnetar formation hypothesis of \citet{duncan92}. As already pointed out
by them, and investigated in detail by \citet{thompson04} and 
\citet{bucciantini07}, 
a rapidly rotating neutron star may produce very energetic supernovae.
The reason is that a large part of the rotational energy of 
$E_{rot} = 3\times 10^{52} (P/1\ {\rm ms})^{-2}$~erg 
will be transfered to the ejecta in less than a few hundred seconds,
due to the strong magnetic torque exerted by the neutron star.
The association of AXPs/SGRs with SNRs allows us therefore to put constraints
on the initial rotation period, by investigating whether these SNRs are more
energetic than other SNRs \citep[c.f.][]{allen04,arons03}.

\subsection{Supernova remnant energies and magnetar initial spin periods}

\citet{vink06c} investigated whether SNRs containing AXPs/SGRs are more
energetic by compiling and/or determining the energies
of the SNRs Kes 73 (AXP 1E 1841-045), CTB 109 (AXP 1E2259+586) and 
N49 (SGR 0526-66). Not investigated was G29.6+0.1, a very faint SNR associated
with candidate AXP AX J1845-045 \citep{vasisht00}. 
The quality of the archival X-ray data of this
SNR is unfortunately too poor for spectroscopic analysis.

The energies of SNRs can be estimated from their X-ray spectra,
by determining the plasma temperature (an indication for the shock velocity),
emission measure (an indication for the density) in conjunction with
an estimate of the distance. The results are 
summarized Table~\ref{tab-energies}. 

It is clear that the energies of these three SNRs are all
consistent with the canonical supernova energy of 
$10^{51}$~erg (but see section~\ref{sec-CTB109}). 
The method for estimating explosion energies
is well tested, see for example \citet{hughes98}, who found explosion 
energies for other SNRs that are comparable to the energies listed
in Table~\ref{tab-energies}.
The method assumes that the SNRs are in the Sedov phase of their
evolution, which
strictly speaking only applies to older SNRs. However, for the young SNR
Kes 73 \citet{vink06c} also considered the \citet{truelove99} model for
young SNRs, which  confirm the relatively low energy of Kes 73.

The implication is that the birth of magnetars cannot have resulted in
an additional energy input of more than $\sim 10^{51}$~erg, which corresponds
to $P_i \ga 5$~ms. This is therefore evidence against the hypothesis
that magnetars are formed with initial periods of $P_i \lesssim 3$~ms.
However, this conclusion is only valid
if most of the rotational energy ended up as kinetic energy of the ejecta.
Moreover, it may also indicate that the convective dynamo is still
effective for initial periods slightly longer than 3~ms.

\begin{table*}
\begin{minipage}{0.92\textwidth}
\caption{The explosion energies and ages of the supernova remnants 
from X-ray spectral analysis (after \citet*{vink06c},
the pulsar dipole field has been taken from \citet{woods04}). 
}
\label{tab-energies}
{\scriptsize
\begin{tabular}{l c c c c c c c c l}\hline\hline\noalign{\smallskip}
SNR/Pulsar & Distance & radius & E &  $n_{\rm H}$ & Mass & SNR Age & 
\ Pulsar Ageb & \ $B_d$\ &References\\
    &kpc   & pc  & ($10^{51}$~erg) & cm$^{-3}$ & \msun & $10^3$yr & $10^3$ yr &
$10^{14}$~G
\\
\noalign{\smallskip} \hline\noalign{\smallskip}
Kes 73/1E1841-045 & 
$7.0$& 4.3 & $0.5\pm0.3$ & $2.9\pm0.4$ & $29\pm4$ & $1.3\pm0.2$&  4.3 & 7.1 &\cite{vink06c}\\
CTB109/1E2259+586 &
$3.0$& 10 & $0.7\pm0.3$ & $0.16\pm0.02$ & $97\pm23$ & $8.8\pm0.9$&  220 & 0.6 &\cite{sasaki04}\\
N49/SGR 0526-66 & 
$50$& 9.3 & $1.3\pm0.3$ & $2.8\pm0.1$ & $320\pm50$ & $6.3\pm$1.0&  1.9 & 7.4 &\cite{vink06c}\\
\noalign{\smallskip}\hline
\end{tabular}\\
}
\medskip
Distances and pulsar ages ($\tau = \frac{1}{2}P/\dot{P}$) are taken from 
\citet{woods04}.
\end{minipage}
\end{table*}

\subsection{The distance to CTB 109}
[\label{sec-CTB109}
It must be pointed out that the estimated explosion energies scale with the 
distance, as $d^{2.5}$. Very recently, \citet{durant06} showed that the
distance to CTB 109 may be twice as large, placing AXP 1E2259+586 in the
Outer Spiral Arm ($d\sim7.5$~kpc) instead of in the Perseus Arm
($d \sim 3$ kpc), and suggesting an explosion
energy of $7\times10^{51}$~erg for CTB 109. 
The distance measurement
consists of comparing the optical absorption of nearby field stars with the
X-ray absorption of the AXP. For this the correlation between 
$N_{\rm H}$\ versus $A_V$\ is used of \citet{predehl95}. 

Although the results
of \citet{durant06} are interesting, the new distance measurement 
should be used cautiously. First of all, there is an intrinsic spread in the
$N_{\rm H}$-$A_V$\ relation. Secondly, \citet{durant06} use the 
$N_{\rm H}$\ measured from the AXP, which is much higher than that derived
for the SNR 
($1.1\times 10^{22}$~cm$^{-2}$ versus $0.7\times 10^{22}$~cm$^{-2}$). 
Adopting the latter would result in a distance
consistent with the previously derived distance of
$\sim 3$~kpc \citep{kothes02}.

The likely reason for the discrepancy between the AXP's and SNR's absorption
is a large variation of absorption columns in this region. Indeed, CTB 109
seems to be interacting with a molecular cloud \citep{sasaki04} and it may
be that this molecular cloud is partly in front of the AXP.
The nearby SNR Cas A has a similarly high variation in absorption. 
Interestingly, Cas A is only 2 degrees away from CTB 109 and 
closer to the Galactic plane. Its distance is 3.4~kpc \citep{reed95},
which places it at the far-side of the Perseus Arm. However, its
absorption column is larger than that of 1E2259+586, namely
$N_H = 1.3\times 10^{22}$~cm$^{-2}$.
This makes the new distance estimate for CTB 109 all the more
peculiar.

The case for a larger distance to CTB 109 is, in my view, not
yet compelling, but it is important to investigate the distance with
different methods:
An energy of $10^{51}$~erg suggests $P_i \lesssim 5$~ms, 
whereas $7\times10^{51}$~erg
is consistent with $P_i < 3$~ms the limit at which significant magnetic field
amplification in the proto-neutron star can take place \citep{duncan96}.

\begin{figure*}
\centerline{
\psfig{figure=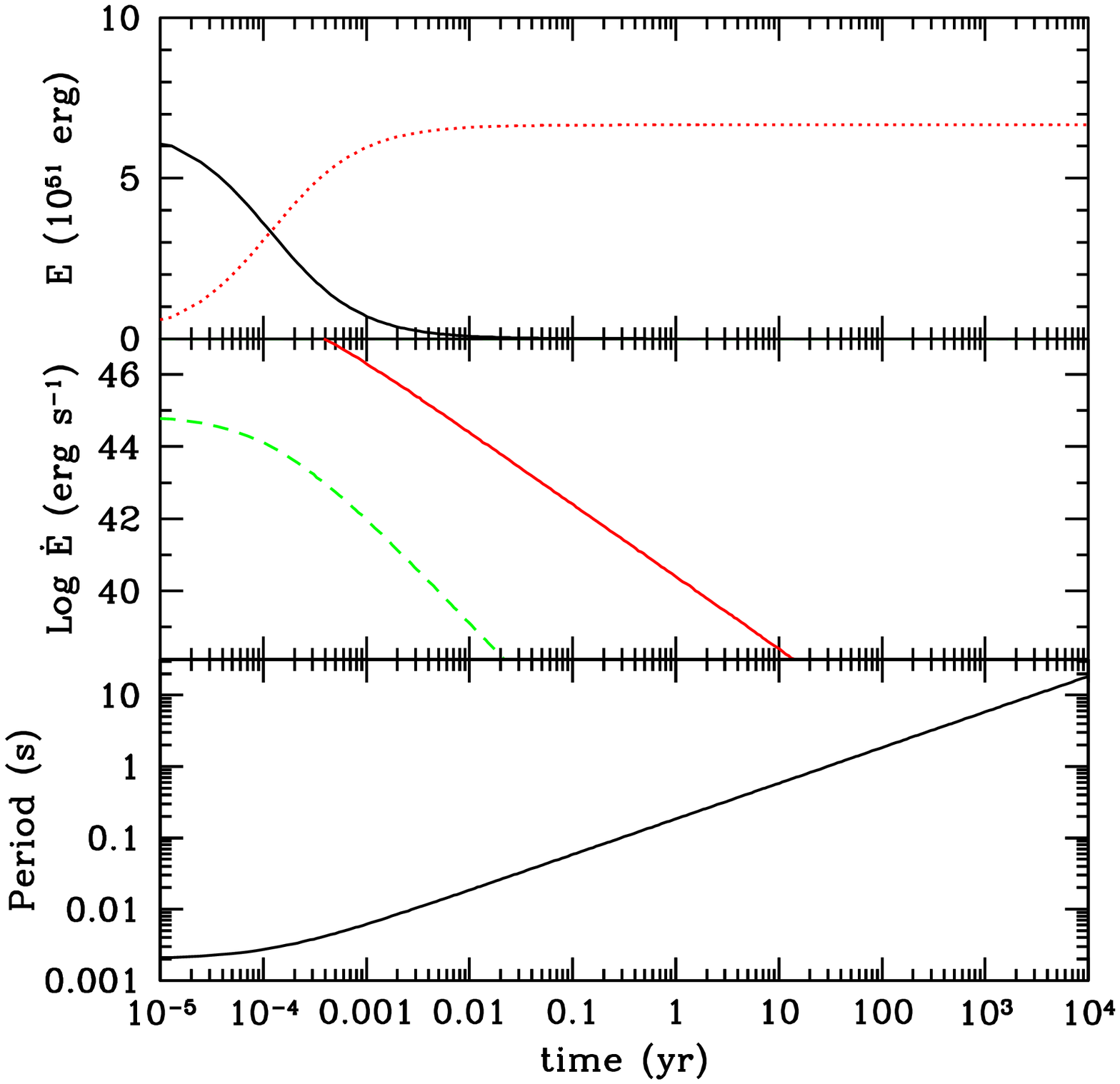,width=0.33\textwidth}
\psfig{figure=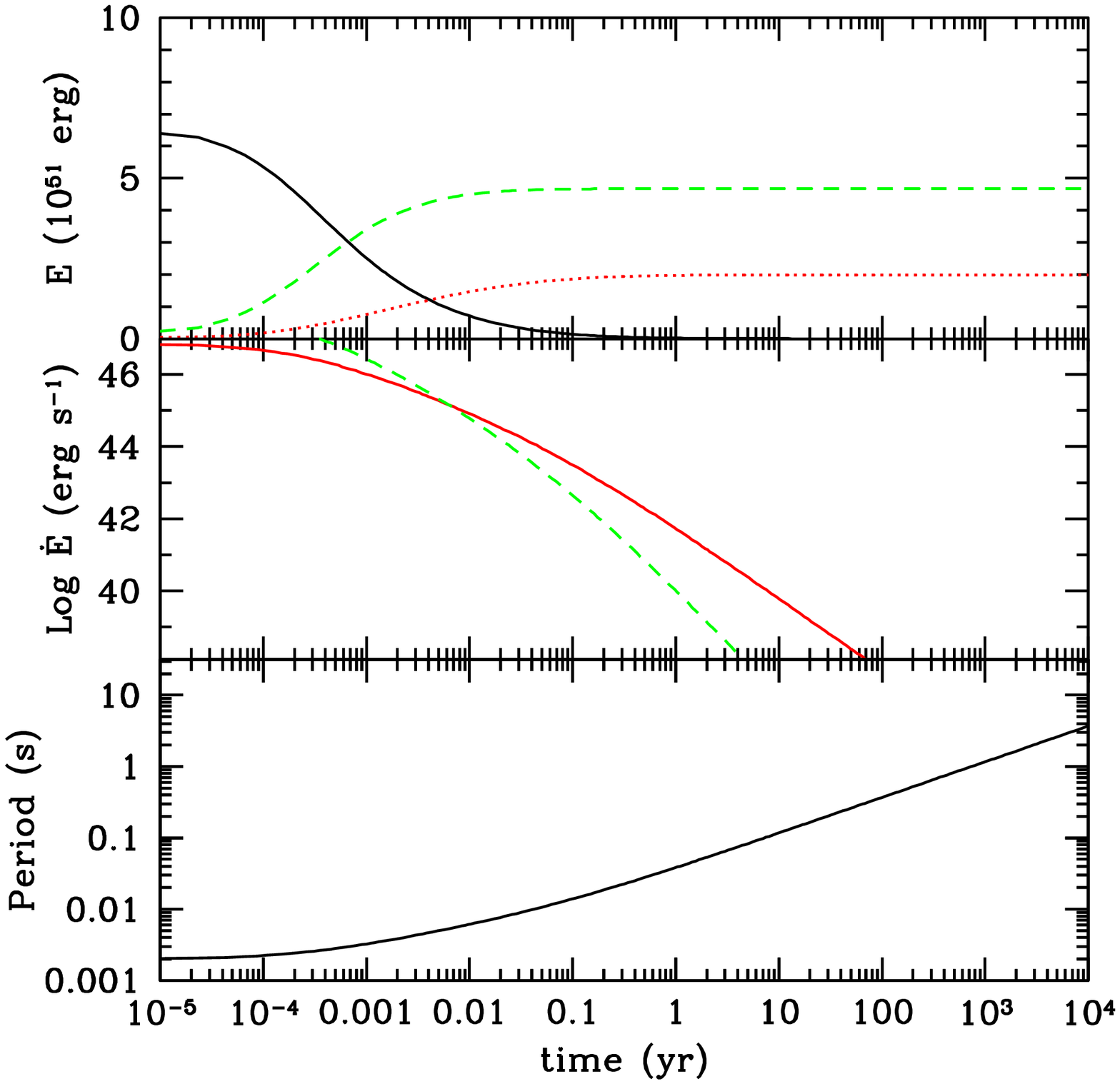,width=0.33\textwidth}
\psfig{figure=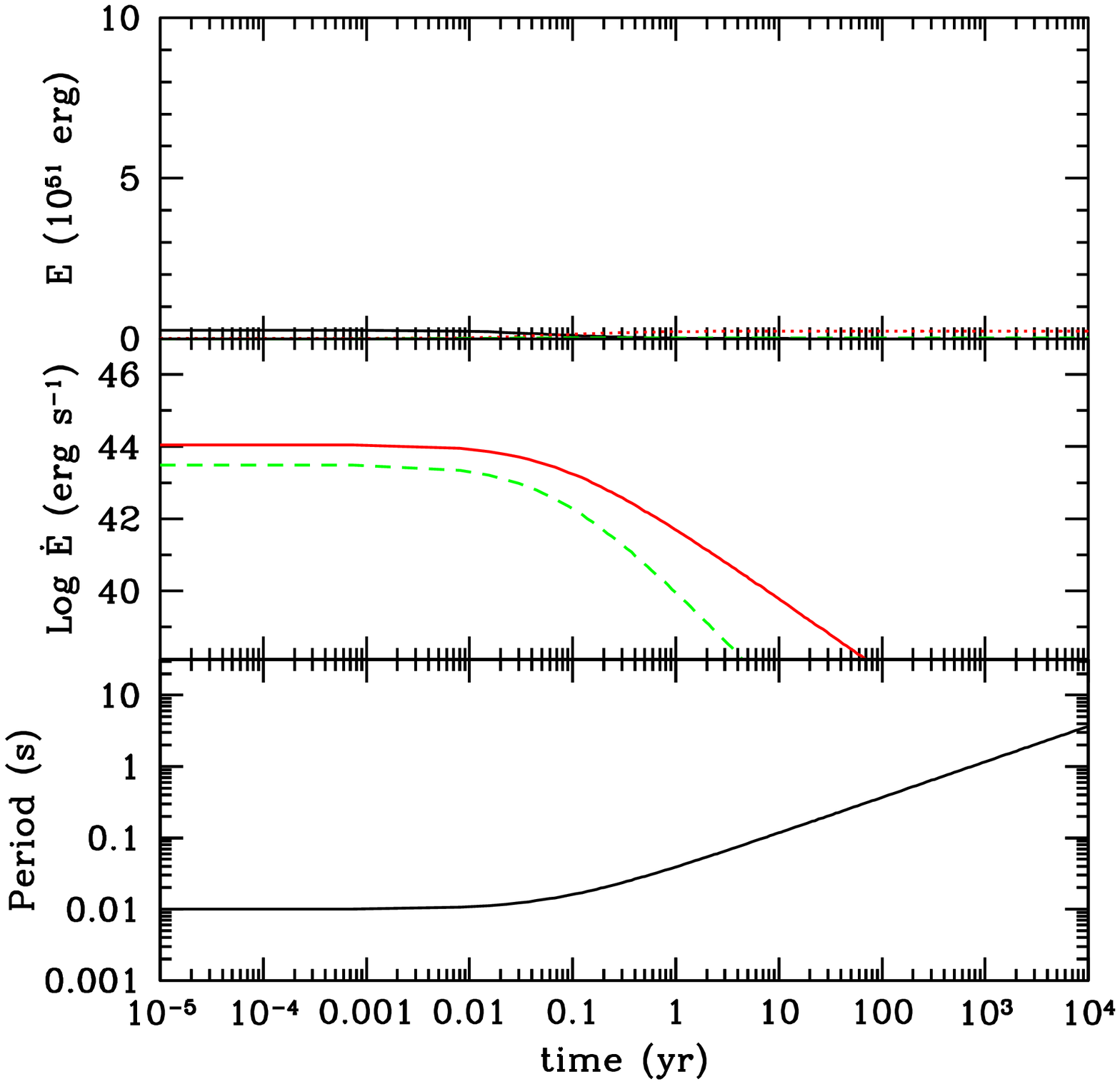,width=0.33\textwidth}
}
\caption{
Simple spindown models for magnetars, assuming the spindown
scenario of \citet{stella05} with different values for the
interior and dipole magnetic field. From left to right:
a) $B_{int}=10^{16}$~G/$B_d=10^{15}$~G, $P_0=2$~ms;
b) $B_{int}=5\times 10^{16}$~G/$B_d=2\times10^{14}$~G, $P_0 = 2$~ms;
c) $B_{int}=10\times 10^{16}$~G/$B_d=2\times 10^{14}$~G, $P_0 = 10$~ms.
The panels show (from top to bottom): 
1) the rotational energy (solid), 
and the integrated rotational energy loss due to
gravitational radiation (dashed) and magnetic braking (dotted);
2) the spindown luminosity in gravitational radiation, and
dipole radiation (for young radio pulsars resulting in a relativistic
wind of similar magnitude); 
3)  the pulse period history.
For comparison a supernova has a peak luminosity of 
$10^{42}-10^{43}$~erg\,s$^{-1}$.
\label{fig-spindown}
}
\end{figure*}
\section{Rotational energy losses}
\label{sec-losses}
Using the SNR energy to constrain the initial spin period of magnetars rests
on the assumption that a large fraction of the rotational energy will 
eventually be dumped into the supernova ejecta \citep[e.g.][]{thompson04}. 
It may be possible that instead rotational energy escapes
in the form of a jet, due to losses in the form of gravitational
radiation, or due to perhaps another yet unknown mechanism.

However, any loss mechanism needs to operate in a time that
is shorter than the magnetic braking time scale, but comparable to, 
or longer than the time scale for magnetic field amplification,
in order not to interfere with the magnetic amplification process:
According to \citet{duncan92} the 
emerging magnetic field damps out differential rotation in less
than 10~s. In vacuum, the magnetic braking time scale is
$\sim 4\times10^3  (P/1\ {\rm ms})^{-2} (B_p/10^{15}\ {\rm G})^{-2}$~s.
However, \citet{thompson04} have shown that for realistic conditions 
the time scale for rotational energy loss may be shorter
than $< 100$~s, and if a relativistic wind emanates from
the proto-neutron star it may even be $<30$~s.

So whatever loss mechanism removes rotational energy without in
the end converting it into kinetic energy, it has to operate on a time scale
between 10~s and less than several hundred seconds.

\subsection{Magnetars and Jets}
Jets are a common phenomenon in almost every accretion process in astrophysics,
from young stellar objects to active galactic nuclei. It is therefore
quite natural to assume that jets may also
form during the core-collapse process.
It has even been argued that jets may be a key ingredient for the supernova
explosion itself \citep[e.g.][]{akayima03}. However, even if a jet forms
it will in most cases not survive the passage through the star.
Exceptions are, of course, supernovae associated with gamma-ray bursts
and X-ray flashes. In fact, it has been argued that the X-ray flash
XRF 060218 was driven by the birth of a magnetar \citep{mazzali06}.

So could it be that magnetar creation is accompanied by jet formation,
which takes away most of the rotational energy, rather than that the
rotational energy drives the ejecta?\footnote{Jet formation may seem  a
natural outcome of accretion processes, but jets are unlikely to provide
an efficient mechanism for angular momentum transport, if aligned with
the angular momentum vector.}

Indeed, it is possible that X-ray flashes
are the result of jet formation associated with
magnetar formation,
as I will discuss in section~\ref{sec-casa}. However,
there is no evidence that the birth of magnetars, in general,
results in the formation of powerful jets:
The morphology of the SNRs listed in Table~\ref{tab-energies}
do not show any evidence for jets.
Kes 73 does not even show signs of asphericity (Fig.~\ref{fig-kes73}).

One may of course wonder whether the explosion was jet induced, but that
by  now the shock has become more spherical.
However, 
this would bring us back to question why SNRs with magnetars have not
more energetic SNR shells.

It is intriguing, though, that the SNR Cas A shows ample evidence
for a bipolar jet structure (see section~\ref{sec-casa}), 
and it has been argued that the
mysterious point source in Cas A is a magnetar:
its X-ray emission properties and lack of detectable radio emission
is reminiscent of AXPs/SGRs \citep{chakrabarty01}, 
and an infrared light echo  of a putative giant SGR flare has been
detected \citep{krause05}.

\subsection{Gravitational waves}

One way for a rapidly spinning neutron star to lose its
rotational energy without powering the supernova ejecta 
is through gravitational radiation.
\citet{stella05} recently showed that magnetar formation
may be accompanied by gravitational radiation, 
provided that the internal magnetic field is unaligned with the rotation axis,
$B_{i} \gtrsim 5\times10^{16}$, and the dipole magnetic field
is relatively low, $B_d < 5\times10^{14}$~G.
The reason is that gravitational radiation is caused by a
deformation of neutron star caused by interior magnetic stresses.
The dipole magnetic field at the surface
determines the magnitude of the competing process, i.e. magnetic braking.
Note, however, that \citet{stella05} assume magnetic braking in vacuum,
which is far from realistic \citep{thompson04}.

Moreover, the dipole surface magnetic field needs to be small; smaller
than the observed magnetic fields of some AXPS/SGRs.
This could mean either that the magnetic field is initially
highly disordered, and dominated by higher order multipoles, or
the magnetic field is buried for some time \citep[e.g.][]{geppert99}.
However, if field emergence is slow, it will be harder
to explain the slow rotational periods of AXPs and SGRs.

The most direct observational test for the idea that magnetar birth
is accompanied by gravitational radiation is to detect gravitational
waves, coinciding with a supernova.
The rotational frequency and its decay with $\dot{\omega} \propto \omega^5$\
should be a clear signature of such an event. As shown by \citet{stella05},
the likelihood for detecting such an event in the Galaxy 
with future gravitational wave detectors is small, but the signal may be strong
enough to detect supernovae as far out as the Virgo cluster.

For the moment we have no means of directly testing whether
magnetars are born with $P < 3$~ms periods and subsequently lose much
of this energy in the form of gravitational waves.
However, even though gravitational radiation losses may be
dominant in the first few minutes, the ``rapidly spinning proto-neutron star''
magnetar formation scenario may still have other observational consequences.

\subsection{Further observational tests for rapid initial rotation}
\label{sec-radioPWN}
Since rotational energy loss by gravitational radiation depends more strongly
on the angular momentum than magnetic braking, sooner or later
magnetic braking ($\dot{\omega} \propto \omega^3$) will dominate
over gravitational wave losses. 
Using the equations used by \citet{stella05}, and assuming an internal field
of $5\times 10^{16}$~G and a dipole field of $10^{14}$~G, one finds that
by the time the period is 12~ms magnetic braking will be the most important
mechanism for angular momentum loss.
Because all pulsars with high spin down luminosities
produce pulsar wind nebula  (PWN),
one may wonder why we do not detect PWNe around magnetars.

The situation of young radio pulsars and magnetars
is not quite comparable: magnetic braking in magnetars is much more rapid.
So the Crab nebula has taken 1000~yr to make the powerful synchrotron
nebula, but a rapidly rotating magnetar will form a relativistic wind bubble
within a year, that is, during the supernova rather than the SNR phase.
In the first few days, instead of creating a PWN,
magnetic braking will probably directly propel
the ejecta \citep[the scenario considered by ][]{vink06c}, 
but as soon as the ejecta have expanded sufficiently,
a relativistic wind will create a bubble of relativistic particles
within the supernova, similar to a PWN.
Depending on the density of the ejecta this PWN may directly heat the ejecta,
causing a prolonged brightness of the supernova, or it may create
a high luminosity PWN, inside the ejecta. 
Again depending on the density of the ejecta,
the UV/X-ray synchrotron radiation may be reprocessed and result in bright
optical/infrared supernovae, 
or it will create a very bright radio and X-ray source. 
The low energy part of the relativistic electron population, 
which  do not suffer as much synchrotron losses, 
may survive for a long time, and may still be visible
as a fossil radio PWN. The presence of the radio PWN may depend on the
strength of the magnetar dipole field in its first ten years.
For very strong fields ($\sim 10^{15}$~G) the pulsar slows down
to $P=0.5$~s during the supernova phase, in which the freely expanding
ejecta are still dense and may influence the shape and evolution of the
(radio) PWN. For $B_d \sim 10^{14}$~G a PWN is still formed after the
supernova phase has seized and a more normal PWN may be formed, whose
radio relic should still be observable.

The enhanced brightness of supernovae is only mildly dependent on the emission 
of strong gravitational radiation, or the exact initial spin period, 
as long as $P_i\lesssim 10$~ms. For strong magnetic dipole fields, 
the supernova's luminosity may not have an enhanced plateau, but will
instead be brighter in the first months. 
One can judge this from the various spin down scenarios
in Fig.~\ref{fig-spindown}, which shows, for relatively modest dipole 
fields, a spin down luminosity of $10^{42}-10^{40}$~erg s$^{-s}$\
in the first 1 to 10 yrs. 

The picture sketched above needs to be modeled more rigorously,
which is well worth the effort, since it may lead to two
observational tests for magnetar formation scenarios,
even for 3~ms $\lesssim P_i \lesssim$ 10~ms:
1) are there supernovae that are exceptionally bright
for more than a year, either in X-rays or in the optical? 2)
are young magnetars surrounded by fossil radio PWN?

None of the known magnetars seem to be surrounded by relic radio PWNe.
The AXP 1E1841-045 lies even within a local minimum of the radio emission
of the SNR Kes 73 \citep{kriss85}.

As far as I know, there is no evidence for supernovae that are unusually
bright for about a year, except perhaps
for SN 1988A in M58, which showed according to \citet{ruiz91} a departure
from the normal radio-active decay powered light curve. However,
this was disputed by \citet{turatto93}. Investigating this issue
in a more systematic way would require long term (1-2 year) monitoring
of nearby supernovae in the optical and X-rays. In essence it provides
a statistical test: given that there are about 220 Galactic core collapse SNRs 
and only 3 are associated
with magnetars one expects that about 1\%\ of the core collapse
supernovae should show the effects of a rapidly rotating magnetar.

Finally, it may be worthwhile to investigate whether radio emission from
supernovae \citep{weiler05} may in some cases be the result of a forming
and expanding PWN, rather than the result of an interaction of the
supernova shock with the progenitor wind. 
From this point of view, the
radio brightening of SN1987J \citep{bietenholz07} is interesting as it may
tell us about the initial spin period of a putative pulsar formed by the 
supernova.

\section{The fossil field hypothesis}

Given the lack of evidence for the rapid spinning scenario for magnetar
formation so far, it is interesting to look into the alternative possibility:
the fossil field hypothesis, 
i.e. the magnetar field is a result
of the high magnetic field in the core of the
progenitor star. The fossil field hypothesis can be divided
in two hypotheses: 1) the strong fossil field hypothesis, which assumes
that the strong progenitors magnetic field is a result of the strong magnetic
field of the cloud from which the star formed 
\citep{ferrario05,ferrario06}, and 
2) the progenitors magnetic field is a result of 
magnetic field amplification inside the progenitor. 

From an observational point of view the (weak and strong) 
fossil field hypothesis is well worth considering. It is well known
that about 5\% of the A stars have a high magnetic field, 1-1000 kG,
constituting the Ap class. Type A stars are  progenitors of
white dwarfs and from a  statistical point of view  
it seems plausible that Ap stars are the progenitors of magnetic white dwarfs
\citep{ferrario05}. 
This also indicates that stellar magnetic fields are
stable enough to survive during the life time of the star.

Observationally it is much harder to determine magnetic fields of O and B 
stars, the progenitors for neutron stars. However, for a few O and B stars
high magnetic fields have been reported.
Interestingly,the B0.2V star $\tau$\,Sco \citep{donati06a} and the
O star HD 191612 \citep{donati06b}, both magnetic stars, seem to have
long rotation periods of 41 and 538 days, respectively. According to
\citet{donati06a}, this
argues in favor of the fossil field hypothesis for reasons already mentioned
in the introduction: if the magnetic field is due to a dynamo process, 
one needs convective motions in conjunction with a short rotation period.

There are, however, still many unsolved questions regarding magnetic fields
in massive stars. First of all, for the fossil field theory it is unclear
how a magnetic field can survive the turbulent formation process of the star.
On the other hand, even the average magnetic field in the interstellar medium
must have an influence on the star forming process, in the sense that
strict flux conservation would result in magnetic fields that provide pressure
support against collapse \citep{mestel56}. Forming stars therefore
need to dissipate or expulse magnetic field energy. A one to one
relation between local interstellar magnetic field and stellar magnetic field
seems therefore rather naive.

Another issue is the stability of the magnetic field. Even if high magnetic
fields are observed in O stars, will this magnetic field survive during the
life time of the star? This may depend on the magnetic field topology;
apparently, some magnetic field configurations, 
a mix of dipolar and toroidal components, are stable against fast decay
\citep{braithwaite04}. Concerning the inevitable slow decay, 
one may speculate that since more massive
stars live shorter, they can end their life with a higher magnetic field.
Within the framework of the fossil field origin for neutron magnetic fields
this may explain observational evidence that magnetars form from the most
massive stars \citep{gaensler05}.

Finally, all measurements of
magnetic field in stars concerns the surface magnetic fields.
However, the neutron star magnetic
field, according to the fossil field theory comes from the stellar core.
The evidence for a connection between magnetic white dwarfs and Ap 
stars gives some reassurance that indeed high magnetic stars have high magnetic
cores, but the evidence is of a statistical nature. For magnetars the fossil
field hypothesis is also hard to prove directly. One potential piece
of evidence would, however, be if a very young magnetar would have
already a long period. One hypothetical case would be if the neutron star
formed in SN1987A turns out to be a slowly rotating neutron star.
However, no neutron star has yet been detected in SN1987A, but interestingly,
there is also no evidence yet of a powerful pulsar \citep{haberl06}.
The current upper limits on the X-ray luminosity from a putative pulsar
are consistent with a bipolar magnetic field of $10^{14}$~G and an initial
period close to $0.5 (B/10^{14})^{1/2}$~s. This should be considered as
an alternative hypothesis next to a more rapidly spinning pulsar with a low
magnetic field, as discussed by \citet{haberl06}.

\section{Was the point source in Cas~A a rapidly spinning proto-neutron star?}
\label{sec-casa}
There are a number of SNRs that contain unresolved X-ray sources,
which are likely to be neutron stars with some unusual properties: 
they show no evidence for
radio emission, and are not surrounded by PWNe. 
They therefore resemble
AXPs, but since no pulsation period has been detected they cannot be positively
identified as magnetars. Moreover, their surface temperatures seem lower
than those of AXPs \citep{pavlov04}. We are hiding our ignorance about these
sources by assigning them to a new class: central compact objects (CCOs).
The fact that CCO's are not surrounded by PWN suggests that their rotational
energy loss is low, meaning that either they have a very
low surface magnetic field, or, perhaps more likely, they are slow rotators.

The canonical CCO is the X-ray point source in Cas A,
detected in the first light image of \chandra\ \citep{tananbaum99}.
Recently, the Spitzer infrared observatory found that infrared emission 
from the vicinity of the SNR shows evidence for (super)luminal motion
\citep{krause05}, attributed to a light echo caused by a luminous outburst
of the CCO in the 1950's. This would imply that the point source is indeed
a magnetar. 

What makes the point source in Cas A equally fascinating is that
it may be a magnetar in a SNR, which also shows the presence
of jets \citep{vink04a,hwang04,hines04,fesen06}.
The jets have an energy of $\sim 10^{50}$~erg \citep[][see also Schure et al. 2007, in preparation]{laming06},
whereas the total supernova explosion energy was probably
$2\times 10^{51}$~erg \citep{laming03}. The explosion
energy, the ejecta mass \citep[2-4~\msun,][]{vink96,willingale02} and the jet 
energies are remarkably similar to the properties derived
for the X-ray flash associated with SN 2006aj \citep{mazzali06}.

Usually it is assumed that gamma-ray bursts are formed from
the collapse of a rapidly rotating stellar core into a black hole 
\citep[collapsar theory][]{macfadyen01},
whereas their weaker counterparts, the X-ray flashes, may be similar, but
instead a rapidly spinning 
neutron star is formed. So how does this relate to the fact that
the point source in Cas A is likely to be a slowly rotating neutron star?
The answer may be that the neutron star has slowed down
considerably in the 330~yrs of its existence, 
as a result of magnetic braking.

If the jets obtained their energy of $\sim 10^{50}$~erg
from the rotation of the neutron star, as argued by
\citet{wheeler02}, this would imply an initial period
of at least $\sim 17$~ms. 
In order to calculate a lower limit to the bipolar magnetic field we need
to estimate from the non-detection of a PWN a lower limit to the current
pulsar period. We can do this as follows. The rotational energy
of a neutron star is $E_{rot}=\frac{1}{2}I \Omega^2$, 
with $I=1.4\times 10^{45}$~g cm$^2$,  thus 
\begin{equation}
\dot{E}_{rot} = I \Omega \dot{\Omega} = 
2 E_{rot} \frac{\dot P}{P} = \frac{E_{rot}}{\tau},\label{eq-edot}
\end{equation}
with $\tau = P/(2 \dot P)$, 
the characteristic pulsar age \citep{seward88}. 
For Cas A we can set $\tau = 330$~yr,
thereby assuming that the neutron star was born with 
$P_i \ll P(t=330\ {\rm yr})$. Most of the visible energy losses of the PWN will
occur in X-rays, with typically $L_X \approx \dot E/100$.
Despite deep X-ray images of Cas A
\citep[e.g.][]{hwang04} no PWN has been detected, so we may safely say that
the X-ray continuum luminosity in the 4-6 continuum band
is from the remnant shell, as shown by the X-ray images 
\citep[e.g.][]{vink03a}. 
The X-ray continuum above 1~keV is approximately a power law up
to energies of $\sim 80$~keV with
$L_X (1-80 {\rm keV}) 
= 4\times 10^{36}$~erg s$^{-1}$ \citep[e.g.][]{vink03a,renaud06}.
Consequently,
the PWN must have a total luminosity substantially less than that.
If we conservatively say that the PWN has $L_X < 10^{36}$~erg s$^{-1}$ (25\%
of the overall X-ray continuum), we obtain according to
Eq.~\ref{eq-edot} $\dot E <  1\times 10^{38}$~erg s$^{-1}$,
and $E_{rot} < 1.0\times 10^{48}$~erg, corresponding 
to a current pulsar period of $P > 160$~ms.\footnote{\citet{seward88} 
give a less conservative lower limit 
of $P > 330$~ms, and $B > 7\times 10^{13}$~G.}

For a neutron star to slow down from a period of $\sim 17$~ms to a period of
$>160$~ms in 330~yr a bipolar magnetic field is necessary of 
$>5\times 10^{13}$~G, suggesting magnetic fields of magnetar strength, 
or a bit less. This extremely interesting given the recent suggestions
that X-ray flashes are powered by newly born magnetars 
\citep{mazzali06,soderberg06}. 
One problem, 
has to be solved in that case:
if the pulsar once had a much higher
rotational energy loss, why don't we observe a fossil radio PWN? 
As discussed in  section~\ref{sec-radioPWN} the answer
may be that the magnetic field of the magnetar is $\sim 10^{15}$~G,
the PWN will then form inside the dense, freely expanding ejecta during
the first ten years of its life.

\section{Summary and conclusions}

Magnetars are one of the hottest topics (literally!) in neutron star research.
It is usually assumed that they are formed from rapidly rotating proto-neutron
stars, and that their magnetic field is the result of a dynamo acting
in the first few seconds of the neutron stars life \citep{duncan92}.

However, as I have discussed, there is not yet any 
observational evidence for this:
the SNRs containing magnetars are not more energetic than other SNRs, excluding
a very rapid initial rotation \citep[$P \gtrsim 5$~ms,][]{vink06c}.
Moreover, the fossil field theory offers a compelling alternative, since
we know that magnetic O and B stars exist, and it is also likely, given the
plausible connection between magnetic A stars and magnetic white dwarfs.

Nevertheless, at this point nothing can be stated with any certainty about 
magnetar magnetic field creation.
It could be that magnetar fields are formed by a convective dynamo,
 even if initial rotations are slightly longer than 3~ms, 
or part of the rotational energy is lost 
in the form of gravitational radiation.

However, I have pointed that the possibilities for testing magnetar formation
scenarios has not yet been exhausted. One might look for the energy input from
a rapidly rotating magnetar in supernovae, which should occur in about 1\% of
the observed core collapse supernovae, or one might look for (faint) relic
radio PWN surrounding the known magnetars.

The fossil field hypothesis is more difficult to directly test, but one
possibility would be to find a slowly rotating magnetar in a very young SNR. 
A possibility
is that the neutron star formed during the SN1987A explosion may turn out
to be a slowly rotating magnetar ($P \gtrsim 0.5$~s). 

Finally, I have pointed out that Cas A may be an
X-ray flash remnant. Since jet formation likely requires
a rapidly rotating compact object \citep{wheeler02},
the X-ray point source
in Cas A may once have been a rapid rotator ($P \lesssim 17$~ms). 
The present day 
lack of a detectable X-ray pulsar wind nebula, implying $P > 160$~ms,
would therefore require it to be a magnetar-like magnetic field 
$B > 5\times 10^{13}$~G.
In order to avoid the creation of a relic radio PWN an even higher
(initial) magnetic field needs to be inferred of $B~\sim 10^{15}$~G.


\begin{thebibliography}{65}
\expandafter\ifx\csname natexlab\endcsname\relax\def\natexlab#1{#1}\fi
\expandafter\ifx\csname url\endcsname\relax
  \def\url#1{\texttt{#1}}\fi
\expandafter\ifx\csname urlprefix\endcsname\relax\def\urlprefix{URL }\fi

\bibitem[{{Akiyama} et~al.(2003){Akiyama}, {Wheeler}, {Meier}, and
  {Lichtenstadt}}]{akayima03}
{Akiyama}, S., {Wheeler}, J.~C., {Meier}, D.~L., {Lichtenstadt}, I., Feb. 2003.
  {The Magnetorotational Instability in Core-Collapse Supernova Explosions}.
  \apj 584, 954--970.

\bibitem[{{Allen} and {Horvath}(2004)}]{allen04}
{Allen}, M.~P., {Horvath}, J.~E., Nov. 2004. {Influence of an Internal Magnetar
  on Supernova Remnant Expansion}. \apj 616, 346--356.

\bibitem[{{Alpar}(2001)}]{alpar01}
{Alpar}, M.~A., Jun. 2001. {On Young Neutron Stars as Propellers and Accretors
  with Conventional Magnetic Fields}. \apj 554, 1245--1254.

\bibitem[{{Arons}(2003)}]{arons03}
{Arons}, J., Jun. 2003. {Magnetars in the Metagalaxy: An Origin for
  Ultra-High-Energy Cosmic Rays in the Nearby Universe}. \apj 589, 871--892.

\bibitem[{{Bhattacharya} and {Soni}(2007)}]{bhattacharya07}
{Bhattacharya}, D., {Soni}, V., May 2007. {A Natural Explanation for
  Magnetars}. ArXiv e-prints 705.

\bibitem[{Bietenholz and Bartel(2007)}]{bietenholz07}
Bietenholz, M., Bartel, N., 2007. {The Evolution of the Central Component in
  SN1986J}. \adspr\ (this volume).

\bibitem[{{Braithwaite} and {Spruit}(2004)}]{braithwaite04}
{Braithwaite}, J., {Spruit}, H.~C., Oct. 2004. {A fossil origin for the
  magnetic field in A stars and white dwarfs}. \nat 431, 819--821.

\bibitem[{{Bucciantini} et~al.(2007){Bucciantini}, {Quataert}, {Arons},
  {Metzger}, and {Thompson}}]{bucciantini07}
{Bucciantini}, N., {Quataert}, E., {Arons}, J., {Metzger}, B.~D., {Thompson},
  T.~A., May 2007. {Magnetar Driven Bubbles and the Origin of Collimated
  Outflows in Gamma-ray Bursts}. ArXiv e-prints 705.

\bibitem[{{Chakrabarty} et~al.(2001){Chakrabarty}, {Pivovaroff}, {Hernquist},
  {Heyl}, and {Narayan}}]{chakrabarty01}
{Chakrabarty}, D., {Pivovaroff}, M.~J., {Hernquist}, L.~E., {Heyl}, J.~S.,
  {Narayan}, R., Feb. 2001. {The Central X-Ray Point Source in Cassiopeia A}.
  \apj 548, 800--810.

\bibitem[{{Chatterjee} and {Hernquist}(2000)}]{chatterjee00}
{Chatterjee}, P., {Hernquist}, L., Nov. 2000. {The Spin Period, Luminosity, and
  Age Distributions of Anomalous X-Ray Pulsars}. \apj 543, 368--372.

\bibitem[{{Donati} et~al.(2006{\natexlab{a}})}]{donati06b}
{Donati}, J.-F., et~al., Jan. 2006{\natexlab{a}}. {Discovery of a strong
  magnetic field on the O star HD 191612: new clues to the future of
  ${\theta}^{1}$ Orionis C$^{*}$}. \mnras 365, L6--L10.

\bibitem[{{Donati} et~al.(2006{\natexlab{b}})}]{donati06a}
{Donati}, J.-F., et~al., Aug. 2006{\natexlab{b}}. {The surprising magnetic
  topology of {$\tau$} Sco: fossil remnant or dynamo output?} \mnras 370,
  629--644.

\bibitem[{{Duncan} and {Thompson}(1992)}]{duncan92}
{Duncan}, R.~C., {Thompson}, C., Jun. 1992. {Formation of very strongly
  magnetized neutron stars - Implications for gamma-ray bursts}. \apjl 392,
  L9--L13.

\bibitem[{{Duncan} and {Thompson}(1996)}]{duncan96}
{Duncan}, R.~C., {Thompson}, C., 1996. {Magnetars}. In: {Rothschild}, R.~E.,
  {Lingenfelter}, R.~E. (Eds.), AIP Conf. Proc. 366: High Velocity Neutron
  Stars. pp. 111--+.

\bibitem[{{Durant} and {van Kerkwijk}(2006)}]{durant06}
{Durant}, M., {van Kerkwijk}, M.~H., Jun. 2006. {Distances to Anomalous X-ray
  Pulsars using Red Clump Stars}. ArXiv Astrophysics e-prints.

\bibitem[{{Ertan} et~al.(2007){Ertan}, {Erkut}, {Ek{\c s}i}, and
  {Alpar}}]{ertan07}
{Ertan}, {\"U}., {Erkut}, M.~H., {Ek{\c s}i}, K.~Y., {Alpar}, M.~A., Mar. 2007.
  {The Anomalous X-Ray Pulsar 4U 0142+61: A Neutron Star with a Gaseous
  Fallback Disk}. \apj 657, 441--447.

\bibitem[{{Ferrario} and {Wickramasinghe}(2006)}]{ferrario06}
{Ferrario}, L., {Wickramasinghe}, D., Mar. 2006. {Modelling of isolated radio
  pulsars and magnetars on the fossil field hypothesis}. \mnras, 292--+.

\bibitem[{{Ferrario} and {Wickramasinghe}(2005)}]{ferrario05}
{Ferrario}, L., {Wickramasinghe}, D.~T., Jan. 2005. {Magnetic fields and
  rotation in white dwarfs and neutron stars}. \mnras 356, 615--620.

\bibitem[{{Fesen} et~al.(2006)}]{fesen06}
{Fesen}, R.~A., et~al., Jul. 2006. {The Expansion Asymmetry and Age of the
  Cassiopeia A Supernova Remnant}. \apj 645, 283--292.

\bibitem[{{Gaensler}(2004)}]{gaensler04}
{Gaensler}, B.~M., 2004. {Anomalous X-ray pulsars and soft gamma-ray repeaters
  - the connection with supernova remnants}. Advances in Space Research 33,
  645--653.

\bibitem[{{Gaensler} et~al.(2005){Gaensler}, {McClure-Griffiths}, {Oey},
  {Haverkorn}, {Dickey}, and {Green}}]{gaensler05}
{Gaensler}, B.~M., {McClure-Griffiths}, N.~M., {Oey}, M.~S., {Haverkorn}, M.,
  {Dickey}, J.~M., {Green}, A.~J., Feb. 2005. {A Stellar Wind Bubble Coincident
  with the Anomalous X-Ray Pulsar 1E 1048.1-5937: Are Magnetars Formed from
  Massive Progenitors?} \apjl 620, L95--L98.

\bibitem[{{Geppert} et~al.(1999){Geppert}, {Page}, and {Zannias}}]{geppert99}
{Geppert}, U., {Page}, D., {Zannias}, T., May 1999. {Submergence and
  re-diffusion of the neutron star magnetic field after the supernova}. \aap
  345, 847--854.

\bibitem[{{Geppert} and {Rheinhardt}(2006)}]{geppert06}
{Geppert}, U., {Rheinhardt}, M., Sep. 2006. {Magnetars versus radio pulsars.
  MHD stability in newborn highly magnetized neutron stars}. \aap 456,
  639--649.

\bibitem[{{Haberl} et~al.(2006){Haberl}, {Geppert}, {Aschenbach}, and
  {Hasinger}}]{haberl06}
{Haberl}, F., {Geppert}, U., {Aschenbach}, B., {Hasinger}, G., Dec. 2006.
  {XMM-Newton observations of <ASTROBJ>SN 1987 A</ASTROBJ>}. \aap 460,
  811--819.

\bibitem[{{Heger} et~al.(2005){Heger}, {Woosley}, and {Spruit}}]{heger05}
{Heger}, A., {Woosley}, S.~E., {Spruit}, H.~C., Jun. 2005. {Presupernova
  Evolution of Differentially Rotating Massive Stars Including Magnetic
  Fields}. \apj 626, 350--363.

\bibitem[{{Hines} et~al.(2004)}]{hines04}
{Hines}, D.~C., et~al., Sep. 2004. {Imaging of the Supernova Remnant Cassiopeia
  A with the Multiband Imaging Photometer for Spitzer (MIPS)}. \apjs 154,
  290--295.

\bibitem[{{Hughes} et~al.(1998){Hughes}, {Hayashi}, and {Koyama}}]{hughes98}
{Hughes}, J.~P., {Hayashi}, I., {Koyama}, K., Oct. 1998. {ASCA X-Ray
  Spectroscopy of Large Magellanic Cloud Supernova Remnants and the Metal
  Abundances of the Large Magellanic Cloud}. \apj 505, 732--748.

\bibitem[{{Hwang} et~al.(2004)}]{hwang04}
{Hwang}, U., et~al., Nov. 2004. {A Million Second Chandra View of Cassiopeia
  A}. \apjl 615, L117--L120.

\bibitem[{{Kaastra} and {Mewe}(2000)}]{kaastra00}
{Kaastra}, J.~S., {Mewe}, R., Oct. 2000. {Coronal Plasmas Modeling and the
  MEKAL code}. In: {Bautista}, M.~A., {Kallman}, T.~R., {Pradhan}, A.~K.
  (Eds.), Atomic Data Needs for X-ray Astronomy. p. 161.

\bibitem[{{Kothes} et~al.(2002){Kothes}, {Uyaniker}, and {Yar}}]{kothes02}
{Kothes}, R., {Uyaniker}, B., {Yar}, A., Sep. 2002. {The Distance to Supernova
  Remnant CTB 109 Deduced from Its Environment}. \apj 576, 169--175.

\bibitem[{{Krause} et~al.(2005)}]{krause05}
{Krause}, O., et~al., Jun. 2005. {Infrared Echoes near the Supernova Remnant
  Cassiopeia A}. Science 308, 1604--1606.

\bibitem[{{Kriss} et~al.(1985){Kriss}, {Becker}, {Helfand}, and
  {Canizares}}]{kriss85}
{Kriss}, G.~A., {Becker}, R.~H., {Helfand}, D.~J., {Canizares}, C.~R., Jan.
  1985. {G27.4+0.0 - A galactic supernova remnant with a central compact
  source}. \apj 288, 703--706.

\bibitem[{{Laming} and {Hwang}(2003)}]{laming03}
{Laming}, J.~M., {Hwang}, U., Nov. 2003. {On the Determination of Ejecta
  Structure and Explosion Asymmetry from the X-ray Knots of Cassiopeia A}. \apj
  597, 347--361.

\bibitem[{{Laming} et~al.(2006){Laming}, {Hwang}, {Radics}, {Lekli}, and
  {Tak{\'a}cs}}]{laming06}
{Laming}, J.~M., {Hwang}, U., {Radics}, B., {Lekli}, G., {Tak{\'a}cs}, E., Jun.
  2006. {The Polar Regions of Cassiopeia A: The Aftermath of a Gamma-Ray
  Burst?} \apj 644, 260--273.

\bibitem[{{MacFadyen} et~al.(2001){MacFadyen}, {Woosley}, and
  {Heger}}]{macfadyen01}
{MacFadyen}, A.~I., {Woosley}, S.~E., {Heger}, A., Mar. 2001. {Supernovae,
  Jets, and Collapsars}. \apj 550, 410--425.

\bibitem[{{Mazzali} et~al.(2006)}]{mazzali06}
{Mazzali}, P.~A., et~al., Aug. 2006. {A neutron-star-driven X-ray flash
  associated with supernova SN 2006aj}. \nat 442, 1018--1020.

\bibitem[{{Mestel} and {Spitzer}(1956)}]{mestel56}
{Mestel}, L., {Spitzer}, Jr., L., 1956. {Star formation in magnetic dust
  clouds}. \mnras 116, 503.

\bibitem[{{Niebergal} et~al.(2006){Niebergal}, {Ouyed}, and
  {Leahy}}]{niebergal06}
{Niebergal}, B., {Ouyed}, R., {Leahy}, D., Jul. 2006. {Magnetic Field Decay and
  Period Evolution of Anomalous X-Ray Pulsarsin the Context of Quark Stars}.
  \apjl 646, L17--L20.

\bibitem[{{Ott} et~al.(2006){Ott}, {Burrows}, {Thompson}, {Livne}, and
  {Walder}}]{ott06}
{Ott}, C.~D., {Burrows}, A., {Thompson}, T.~A., {Livne}, E., {Walder}, R., May
  2006. {The Spin Periods and Rotational Profiles of Neutron Stars at Birth}.
  \apjs 164, 130--155.

\bibitem[{{Ouyed} et~al.(2006){Ouyed}, {Niebergal}, {Dobler}, and
  {Leahy}}]{ouyed06}
{Ouyed}, R., {Niebergal}, B., {Dobler}, W., {Leahy}, D., Dec. 2006.
  {Three-Dimensional Simulations of the Reorganization of a Quark Star's
  Magnetic Field as Induced by the Meissner Effect}. \apj 653, 558--567.

\bibitem[{{Pavlov} et~al.(2004){Pavlov}, {Sanwal}, and {Teter}}]{pavlov04}
{Pavlov}, G.~G., {Sanwal}, D., {Teter}, M.~A., 2004. {Central Compact Objects
  in Supernova Remnants}. In: IAU Symposium. pp. 239.

\bibitem[{{Predehl} and {Schmitt}(1995)}]{predehl95}
{Predehl}, P., {Schmitt}, J.~H.~M.~M., Jan. 1995. {X-raying the interstellar
  medium: ROSAT observations of dust scattering halos.} \aap 293, 889--905.

\bibitem[{{Reed} et~al.(1995){Reed}, {Hester}, {Fabian}, and
  {Winkler}}]{reed95}
{Reed}, J.~E., {Hester}, J.~J., {Fabian}, A.~C., {Winkler}, P.~F., Feb. 1995.
  {The three-dimensional structure of the Cassiopeia A supernova remnant. I.
  The spherical shell}. \apj 440, 706--+.

\bibitem[{{Renaud} et~al.(2006)}]{renaud06}
{Renaud}, M., et~al., Aug. 2006. {The Signature of $^{44}$Ti in Cassiopeia A
  Revealed by IBIS/ISGRI on INTEGRAL}. \apjl 647, L41--L44.

\bibitem[{{Ruiz-Lapuente} et~al.(1991){Ruiz-Lapuente}, {Kidger}, {Gomez},
  {Canal}, and {Lopez}}]{ruiz91}
{Ruiz-Lapuente}, P., {Kidger}, M., {Gomez}, G., {Canal}, R., {Lopez}, R., Sep.
  1991. {SN 1988A in M58 - Departure from Co-56 decay 700 days after
  explosion}. \apjl 378, L41--L44.

\bibitem[{{Sasaki} et~al.(2004){Sasaki}, {Plucinsky}, {Gaetz}, {Smith},
  {Edgar}, and {Slane}}]{sasaki04}
{Sasaki}, M., {Plucinsky}, P.~P., {Gaetz}, T.~J., {Smith}, R.~K., {Edgar},
  R.~J., {Slane}, P.~O., Dec. 2004. {XMM-Newton Observations of the Galactic
  Supernova Remnant CTB 109 (G109.1-1.0)}. \apj 617, 322--338.

\bibitem[{{Seward} and {Wang}(1988)}]{seward88}
{Seward}, F.~D., {Wang}, Z.-R., Sep. 1988. {Pulsars, X-ray synchrotron nebulae,
  and guest stars}. \apj 332, 199--205.

\bibitem[{{Soderberg} et~al.(2006)}]{soderberg06}
{Soderberg}, A.~M., et~al., Aug. 2006. {Relativistic ejecta from X-ray flash
  XRF 060218 and the rate of cosmic explosions}. \nat 442, 1014--1017.

\bibitem[{{Spruit}(2002)}]{spruit02}
{Spruit}, H.~C., Jan. 2002. {Dynamo action by differential rotation in a stably
  stratified stellar interior}. \aap 381, 923--932.

\bibitem[{{Stella} et~al.(2005){Stella}, {Dall'Osso}, {Israel}, and
  {Vecchio}}]{stella05}
{Stella}, L., {Dall'Osso}, S., {Israel}, G.~L., {Vecchio}, A., Dec. 2005.
  {Gravitational Radiation from Newborn Magnetars in the Virgo Cluster}. \apjl
  634, L165--L168.

\bibitem[{{Tananbaum}(1999)}]{tananbaum99}
{Tananbaum}, H., Sep. 1999. {Cassiopeia A}. \iaucirc 7246, 1.

\bibitem[{{Thompson} and {Duncan}(1993)}]{thompson93}
{Thompson}, C., {Duncan}, R.~C., May 1993. {Neutron star dynamos and the
  origins of pulsar magnetism}. \apj 408, 194--217.

\bibitem[{{Thompson} et~al.(2004){Thompson}, {Chang}, and
  {Quataert}}]{thompson04}
{Thompson}, T.~A., {Chang}, P., {Quataert}, E., Aug. 2004. {Magnetar Spin-Down,
  Hyperenergetic Supernovae, and Gamma-Ray Bursts}. \apj 611, 380--393.

\bibitem[{{Truelove} and {McKee}(1999)}]{truelove99}
{Truelove}, J.~K., {McKee}, C.~F., Feb. 1999. {Evolution of Nonradiative
  Supernova Remnants}. \apjs 120, 299--326.

\bibitem[{{Turatto} et~al.(1993){Turatto}, {Cappellaro}, {Benetti}, and
  {Danziger}}]{turatto93}
{Turatto}, M., {Cappellaro}, E., {Benetti}, S., {Danziger}, I.~J., Nov. 1993.
  {Observations of Type-II Plateau Supernovae - Supernova 1988A Supernova 1988H
  and Supernova 1989C}. \mnras 265, 471--+.

\bibitem[{{Vasisht} et~al.(2000){Vasisht}, {Gotthelf}, {Torii}, and
  {Gaensler}}]{vasisht00}
{Vasisht}, G., {Gotthelf}, E.~V., {Torii}, K., {Gaensler}, B.~M., Oct. 2000.
  {Detection of a Compact X-Ray Source in the Supernova Remnant G29.6+0.1: A
  Variable Anomalous X-Ray Pulsar?} \apjl 542, L49--L52.

\bibitem[{{Vink}(2004)}]{vink04a}
{Vink}, J., Feb. 2004. {X- and {$\gamma$}-ray studies of Cas A: exposing core
  collapse to the core}. New Astronomy Review 48, 61--67.

\bibitem[{{Vink} et~al.(1996){Vink}, {Kaastra}, and {Bleeker}}]{vink96}
{Vink}, J., {Kaastra}, J.~S., {Bleeker}, J.~A.~M., Mar. 1996. {A new mass
  estimate and puzzling abundances of SNR Cassiopeia A.} \aap 307, L41--L44.

\bibitem[{{Vink} and {Kuiper}(2006)}]{vink06c}
{Vink}, J., {Kuiper}, L., Jul. 2006. {Supernova remnant energetics and
  magnetars: no evidence in favour of millisecond proto-neutron stars}. \mnras
  370, L14--L18.

\bibitem[{{Vink} and {Laming}(2003)}]{vink03a}
{Vink}, J., {Laming}, J.~M., Feb. 2003. {On the magnetic fields and particle
  acceleration in Cassiopeia A}. \apj 584, 758--769.

\bibitem[{{Weiler} et~al.(2005){Weiler}, {van Dyk}, {Sramek}, {Panagia},
  {Stockdale}, and {Montes}}]{weiler05}
{Weiler}, K.~W., {van Dyk}, S.~D., {Sramek}, R.~A., {Panagia}, N., {Stockdale},
  C.~J., {Montes}, M.~J., Dec. 2005. {Radio Emission from Supernovae}. In:
  {Turatto}, M., {Benetti}, S., {Zampieri}, L., {Shea}, W. (Eds.), ASP Conf.
  Ser. 342: 1604-2004: Supernovae as Cosmological Lighthouses. pp. 290--+.

\bibitem[{{Wheeler} et~al.(2002)}]{wheeler02}
{Wheeler}, J.~C., et~al., Apr. 2002. {Asymmetric Supernovae from
  Magnetocentrifugal Jets}. \apj 568, 807--819.

\bibitem[{{Willingale} et~al.(2002){Willingale}, {Bleeker}, {van der Heyden},
  {Kaastra}, and {Vink}}]{willingale02}
{Willingale}, R., {Bleeker}, J.~A.~M., {van der Heyden}, K.~J., {Kaastra},
  J.~S., {Vink}, J., Jan. 2002. {X-ray spectral imaging and Doppler mapping of
  Cassiopeia A}. \aap 381, 1039--1048.

\bibitem[{{Woltjer}(1964)}]{woltjer64}
{Woltjer}, L., Oct. 1964. {X-Rays and Type I Supernova Remnants.} \apj 140,
  1309--1313.

\bibitem[{{Woods} and {Thompson}(2004)}]{woods04}
{Woods}, P., {Thompson}, C., 2004. Soft gamma repeaters and anomalous x-ray
  pulsars: Magnetar candidates. in Compact Stellar X-ray Sources, eds. W.H.G.
  Lewin and M. van der Klis (astro-ph/0406133).

\end{thebibliography}
\end{document}